\documentclass[11pt,a4paper]{article}
\usepackage[hyperref]{acl2020}
\usepackage{times}
\usepackage{latexsym}

\usepackage{microtype}
\usepackage{array}
\usepackage{balance}
\usepackage{subfigure}
\usepackage{multirow}
\usepackage{float}
\usepackage{color}
\usepackage{soul}
\usepackage{mdframed}
\usepackage{amsopn}
\usepackage{mathrsfs}
\usepackage{mathtools}
\usepackage{amsmath}
\usepackage{amssymb}
\usepackage{bbm}
\usepackage{arydshln}
\usepackage{hyperref}
\usepackage{multicol}
\usepackage{blkarray}
\usepackage{enumerate}
\usepackage{courier}
\usepackage{rotating}
\usepackage{booktabs}
\usepackage{diagbox}
\usepackage{fancybox}
\usepackage{minibox}
\usepackage{cases}
\usepackage[noend]{algpseudocode}
\usepackage{algorithm}
\usepackage{enumitem}
\usepackage{adjustbox}

\usepackage[font=small,labelfont=bf,tableposition=top]{caption}
\aclfinalcopy

\newcommand{\approach}{\textsc{Cookie}}

\title{\textsc{Cookie}: A Dataset for Conversational Recommendation\\
over Knowledge Graphs in E-commerce}

\author{Zuohui Fu${^\dagger}$, Yikun Xian${^\dagger}$, Yaxin Zhu${^\dagger}$, Yongfeng Zhang${^\dagger}$, Gerard de Melo${^\ddagger}$\\
  ${^\dagger}$Rutgers University, NJ, USA \\
  ${^\ddagger}$HPI/University of Potsdam, Germany \\
  \texttt{\{zuohui.fu,yikun.xian,yaxin.zhu,yongfeng.zhang\}@rutgers.edu}\\
  \texttt{gdm@demelo.org}\\}

\date{}

\begin{document}

\maketitle

\begin{abstract}
In this work, we present a new dataset for conversational recommendation over knowledge graphs in e-commerce platforms called \textsc{Cookie}. The dataset is constructed from an Amazon review corpus by integrating both user--agent dialogue and custom knowledge graphs for recommendation.
Specifically, we first construct a unified knowledge graph and extract key entities between user--product pairs, which serve as the skeleton of a conversation. Then we simulate conversations mirroring the human coarse-to-fine process of choosing preferred items.
The proposed baselines and experiments demonstrate that our dataset is able to provide innovative opportunities for conversational recommendation.
\end{abstract}

\section{Introduction}
The rapid development of conversational systems has had substantial impact in industry, but remains under-explored in e-commerce settings.
When choosing products or services, customers may easily feel overwhelmed or confused by the various technical specs and product details \cite{Bettman1998}.
Recently, conversational recommender systems have been proposed to interactively and dynamically solicit information about user requirements so as to provide better recommendations \cite{jannach2020survey}.
\begin{figure}[t]
\centering
\includegraphics[width=\linewidth]{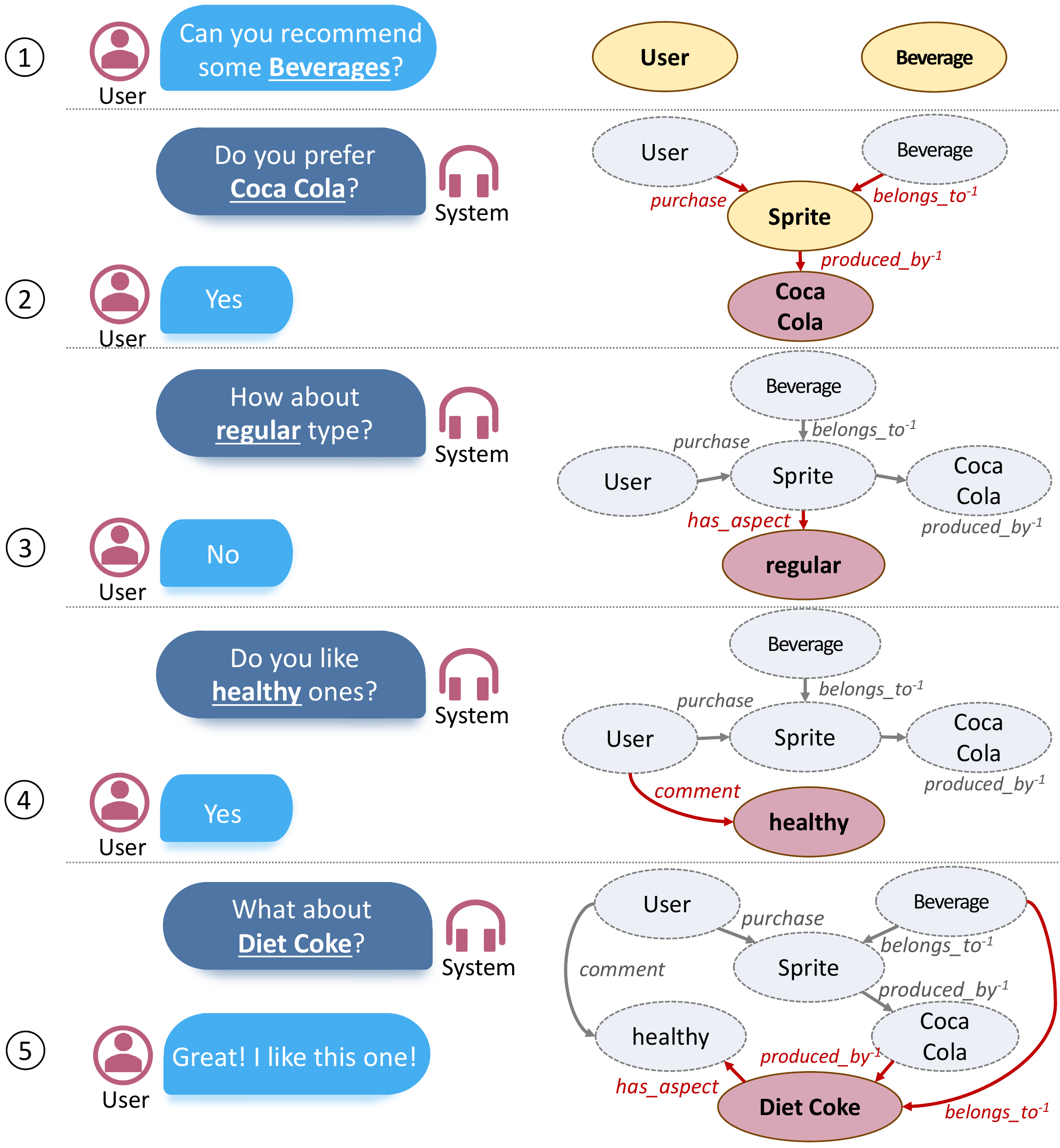}
\caption{A knowledge-enhanced conversational recommender system aims to interact with the user to predict user preferences and make recommendations.}
\label{fig:motivation}

\end{figure}
At the same time, knowledge graphs have recently come to prominence to endow recommender systems with explainability and transparency  \cite{Zhang2018ExplainableRA}, as their graph structure makes it easy to trace connections from users to specific recommendations, and the discovered paths can be presented to the customer.
However,
previous work
neglects user-side information as a part of the graph, only enriching product information with external knowledge bases such as Freebase~\cite{zhao2019kb4rec}.
Thus, it is promising to integrate user activities as well as historical user preferences into the knowledge graph such that the conversational system can better assist users to effortlessly find the best-suited products.

In practice, humans often proceed in a coarse-to-fine manner to gradually make their decisions. For example, people answer questions by first skimming the text, identifying key ideas, and then carefully reading specific parts to obtain an answer~\cite{Masson1983ConceptualPO}.
Similarly, customers often initiate queries to an e-commerce conversational engine that describe the sought products in broader terms, e.g., categories or brand names. During the interaction with the system, the latter gradually gains a better understanding of specific user requirements and preferences pertaining to the relevant products to be chosen.
Hence, the success of such a conversation hinges on the richness of the acquired knowledge by the system.
In order to make the user--agent conversations more reasonable and transparent, we draw on a unified knowledge graph based on the Amazon review corpus by \newcite{ni2019justifying}.
Specifically, for every user, we determine the set of reachable entities connected with purchased products as essential sequential knowledge within the KG, aiming to follow natural
coarse-to-fine conceptual resolution to gradually propagate the user interests.

To this end, we present a novel corpus called \approach{}:
\textbf{CO}nversational recommendation \textbf{O}ver \textbf{K}nowledge graphs \textbf{I}n \textbf{E}-commerce platforms.
The key contributions of this paper can be summarized as follows.
1) We highlight the importance of injecting unified knowledge graphs into conversational recommendation, and induce a corresponding dataset to encourage further research.
2) We propose a simple yet effective pipeline to construct a knowledge graph based on the Amazon review dataset and identify key entities that can be invoked to generate conversations.
3) We provide several baseline results for recommendation and next-question prediction.

\section{Related Work}
We review several essential features of state-of-art conversational recommendation.
\citet{zhang2018towards} design a multi-memory network architecture, which can ask aspect-based questions to gradually understand the user preferences.
However, it fails to consider human-readable utterances as a fluent response for semantic understanding, which is replaced by either extracting the facets from the utterance or crawling the raw review contexts.
In contrast, \citet{li2018conversational} encode a dialogue via an RNN-based neural network to extract the dialogue state.
\citet{Greco2017ConverseEtImperaED} propose a framework based on hierarchical RL for dialogue management. These works focus on dialogue generation and recommendation.
However, retrieval-based conversational engines such as AliMe \cite{Li2017AliMeA} as e-commerce assistants have proven more popular in practice. Compared to generation-based methods \cite{Liu2019RhetoricallyCE, Chen2019TowardsKR}, retrieval-based methods are often able to provide more fluent and informative responses \cite{Yang2018ResponseRW,Yuan2019MultihopSN}.
Recently, the integration of knowledge graphs (KGs) has enabled  recommendation grounded in reasoning in conjunction with conversational knowledge.
\citet{moon2019opendialkg} propose an attention-based graph decoder that seeks optimal paths within a KG, and a zero-shot learning model that leverages previous sentences, dialogue, and KG contexts to re-rank candidates from the pruned decoder graph output.
In \citet{Chen2019KBRD}, item-related knowledge bases with entity-linked text lead to better performance than either of them alone in dialogue generation and recommendation. Comparing to these methods, we provide an open dataset for conversational recommendation that integrates knowledge graphs so that prominent knowledge with semantics can be used to provide both personalized and explainable recommendation.

\section{Dataset Construction and Task}
In this section, we describe the pipeline to construct \approach and the corresponding task definition.
Before that, we first describe the key desiderata.
\textbf{Manually verified}:
Most commercial conversational engines principally rely on template-based utterance generation (e.g., Alexa Skills, DialogueFlow, etc.). This requires substantial development effort, which however is tied to a particular model.
Manually verified data has the advantage of allowing the data construction to be completed separately from model development and learning.
\textbf{Reliability}: Although conversations may be simulated, the generated questions and user responses should be reasonable.
\textbf{Personalization}: One of the cornerstones of recommendation is that the results are personalized, accounting for the specific historical records available for each user. Thus, even for two otherwise identical conversations, we expect diverse recommendation results based on the user's past activities.
\textbf{Goal-Oriented}: Users of e-commerce platforms tend to be impatient and hence the conversation should not be lengthy, as opposed to open-domain chatbot-style dialogue. Rather than getting the user involved in a long conversation spanning many rounds, a key objective is to satisfy the user's needs as efficiently as possible and quickly identify personalized target items.

Existing methods for conversational recommender systems are either based on dialogue state tracking \cite{Sun2018ConversationalRS,Lei2020EstimationActionReflectionTD}, which typically represents the dialogue state by facet attributes of items, or on dialogue semantic modeling \cite{zhang2018towards}, which focuses on understanding the semantics of the dialogue via language models. We draw on the KG structure and on the dialogue and try to unify these two philosophies. The goal is to predict the next utterance while simultaneously addressing next-question prediction as well as the final recommendation task.

The four domains of our dataset are Cellphones \& Accessories, Grocery \& Gourmet, Toys \& Games, and Automotive (see Table~\ref{tab:stats}). Each category is a separate domain of the e-commerce platform and is hence considered as an independent sub-dataset. The pipeline involves first constructing a knowledge graph, followed by the process of key entity extraction and finally conversation synthesis.

\smallskip
\noindent\textbf{Unified Knowledge Graph Construction.}
We start from a recent collection of Amazon reviews \cite{ni2019justifying}.
The extracted facts can mainly be categorized into two groups: user activities and product meta-data.
For user activity related facts, we extract user review keywords and liked styles of products following \newcite{zhang2014explicit,zhang2018towards}. This yields multiple categories of user records (purchases, comments, etc.) and abundant product information (price, aspects, category, brand, etc.).
The unified knowledge graphs in this work not only capture user activities towards products but also incorporate rich product meta-information.

\newcolumntype{R}[1]{>{\raggedleft\let\newline\\\arraybackslash\hspace{0pt}}m{#1}}
\begin{table}[t]
\centering
\scriptsize
\begin{tabular*}{\linewidth}{@{}p{0.9cm}R{1.5cm}R{1.3cm}R{1.14cm}R{1.14cm}@{}}
\toprule
& \textbf{Cellphones \& Accessories} & \textbf{Grocery \& Gourmet} & \textbf{Toys \& Games} & \textbf{Automotive} \\
\midrule
\#Entities  & 278,198   & 271,855    & 437,897   & 444,545   \\
\#Relations & 45        & 45         & 71        & 73        \\
\#Triples   & 3,724,724 & 4,452,234  & 6,705,842 & 5,703,094 \\
\#Interactions  & 607,673   & 709,280    & 1,178,943 & 1,122,776 \\
\#Utterances & 2,043,988 & 2,424,103  & 3,339,771 & 3,830,556 \\
\bottomrule
\end{tabular*}
\vspace{1mm}
\caption{Dataset statistics. \#Entities include total number of users, products and all other KG entities. \#Relations represents the number of unique relation types. \#Triples is the number of triples except for user--product iterations, which is denoted as \#Interactions
and \#Utterances represents the total number of utterances.
}

\label{tab:stats}
\end{table}

\smallskip
\noindent\textbf{Key Entity Extraction.}
Once the KG is constructed, the next step is to consider each ground truth user--product interaction and extract relevant key entities from the knowledge graph that motivate the purchase decision.
In Fig.~\ref{fig:motivation}, for instance, the key entities highlighted in red include product categories, attributes such as \emph{healthy}, etc.
Each sequence of key entities
later serves as a skeleton for the respective dialogue, guiding a coarse-to-fine selection process in which the entities determine which feature is considered in each conversational turn.
Therefore, we sort the entities by node degree, and then select the KG entities that are reachable from the given user and product within one or two hops.
The underlying intuition is that since the conversational system aims to help users to gradually figure out their preferences,
the system starts from larger degree entities, as these are more prominent, well-known, and often more generic. As the conversation proceeds, the latent needs of users are progressively clarified such that it becomes easier to consider key entities with a smaller degree, i.e., more particular fine-grained ones.

\smallskip
\noindent\textbf{Conversation Synthesis.}
The next step is to generate dialogue for the recommendation interactions.
For each ground truth user--product pair, we compose the corresponding conversations based on the skeleton formed by the respective sequence of key entities.
In particular, we transform the key entities into questions via human-specified templates $Q(.)$ generated from \newcite{Wiseman2018LearningNT}, which are manually verified and require simple Yes/No-style answers from the user. Apart from simplifying the dataset creation and subsequent prediction, it also makes sense to assume that those users seeking assistance rather than directly selecting an item tend to be unfamiliar with the product details and are unable to provide detailed requirements. In this case, Yes/No questions are a natural way of narrowing down the search space.

We simulate a conversation procedure in a coarse-to-fine manner to construct the dataset.
Formally, we define a $T$-turn \emph{knowledge-enhanced conversation} as \[C^{(T)}=(q_0, (q_1, a_1, e_1), \ldots, (q_T, a_T, e_T)),\]
where $q_0$ is the query initiated by the user, $q_t~(t=1,\ldots,T)$ is the $t$-th question given by the agent, and $a_t~(t=1,\ldots,T)$ is the $t$-th answer given by the user.
Assume that each question
$q_t$
is associated with an entity $e_t\in\mathcal{E}$, where $\mathcal{E}$ denotes the entity set of an knowledge graph $\mathcal{G}$.
Given $C^{(T)}$, we will expect the model to make two kinds of predictions at step $T+1$: next-question prediction and recommendation. For these, we need the set of candidate questions $Q^{(T+1)}$, candidate key entities $E^{(T+1)}$, and candidate items $V$.
The details for constructing these for our dataset are as follows.
For each user $u_i$ and item $v_j$ purchased by that user, we take as input a sequence of $T+1$ key entities $\{e_0,\ldots,e_T\}$, as obtained in the previous section, along with a sequence of corresponding answers $\{a_1,\ldots,a_T\}$. Here, $e_0$ is a key entity identified from the user query, so there is no corresponding answer for it.
We first construct the $T$-turn conversation $C_{ij}^{(T)}$ for user $u_i$ and item $v_j$ via question templates $Q(\cdot)$ and obtain: $C_{ij}^{(T)}=(q_0, (Q(e_1), a_1, e_1), \ldots, (Q(e_T), a_T, e_T))$.
Then, we build the three candidate sets via negative sampling.
For item set $V_{ij}$ of user $u_i$ and item $v_j$, we randomly sample a subset of $|V|-1$ items that the user has not purchased, denoted by $v_1^-,\ldots,v_{|V|-1}^-$, and derive the candidate item set $V=\{v_j,v_1^-,\ldots,v_{|V|-1}^-\}$.
To construct the candidate entity set $E_{ij}^{(T+1)}$ for the $T+1$-th turn, we first sample a set of paths from the user $u_i$ to item $v_j$ and randomly retrieve $N$ nodes from these paths, denoted by $e_1^-,\ldots,e_{N}^-$. Thus, the candidate entity set can be formed as $E_{ij}^{(T+1)}=\{e_{T+1},e_1^-,\ldots,e_{N}^-\}$, where $e_{T+1}$ is the ground-truth key entity previously obtained.
Accordingly, the candidate question set $Q_{ij}^{(T+1)}$ can be generated from templates and candidate entities, i.e., $Q_{ij}^{(T+1)} = \{Q(e)|e\in E_{ij}^{(T+1)}\}$.
Since we know the ground-truth of the next question $q_{T+1}=Q(e_{T+1})$, the next entity $e_{T+1}$, and the purchased item $v_j$, binary labels can be also provided indicating whether or not a model makes a correct prediction.

\noindent\textbf{Task Formulation.}
The problem of knowledge graph enhanced conversational recommendation is formulated as follows.
Given a $T$-turn knowledge enhanced conversation $C^{(T)}$ and three candidate sets of questions $Q^{(T+1)}$, entities $E^{(t+1)}$, and items $V$, the goal is to predict (i) the next question $q_{T+1}\in Q^{(T+1)}$ in turn $T+1$, (ii) the corresponding key entity $e_{T+1}\in E^{(T+1)}$, and (iii) top $k$ items for recommendation $\{v_{(1)},\ldots,v_{(k)}\}\subseteq V$.

\begin{table}[t]
\centering
\scriptsize
\begin{tabular*}
{\linewidth}{@{}p{1.1cm}p{1.2cm}p{1.2cm}p{1.2cm}p{1.2cm}@{}}
\toprule
& \textbf{Cellphones\& Accessories} & \textbf{Grocery\& Gourmet} & \textbf{Toys\& Games} & \textbf{Automotive} \\
\midrule
 BPR  &0.540  &0.521  &0.498  &0.487 \\
 KGAT &0.593  &0.622  &0.637  &0.581 \\
 OpenDialKG & 0.480  & 0.502 & 0.446  & 0.498\\
 KBRD &0.424  &0.475  &0.366  &0.409 \\
\bottomrule
\end{tabular*}
\vspace{1mm}
\caption{F$_1$@10 results of next-question prediction. Evaluation based on samples of 100 negative products as candidates.}
\label{tab:eval}
\end{table}

\begin{table}[t]
\centering
\scriptsize
\begin{tabular*}{\linewidth}{@{}p{1.1cm}p{1.2cm}p{1.2cm}p{1.2cm}p{1.2cm}@{}}
\toprule
& \textbf{Cellphones\& Accessories} & \textbf{Grocery\& Gourmet} & \textbf{Toys\& Games} & \textbf{Automotive} \\
\midrule
 DMN & 0.414  & 0.429 & 0.392  & 0.388   \\
 DAM & 0.448  & 0.501 & 0.462  & 0.490\\
 MSN & 0.584 & 0.617   & 0.595  & 0.587\\
 OpenDialKG  & 0.670 & 0.710  & 0.535  & 0.707  \\
 KBRD &  0.666 & 0.792& 0.703   & 0.713\\
\bottomrule
\end{tabular*}
\vspace{1mm}
\caption{Recall@2 results on next-question prediction. Evaluation using samples of 100 negative products as candidates.}
\label{tab:res_next}

\end{table}

\section{Baselines and Experiments}
In this section, we evaluate the recommendation and next-question prediction tasks over our constructed conversation dataset, where each sub-dataset is divided into training (60\%), validation (20\%), and test portions (20\%).
In terms of methods, for the recommendation task, we compare Bayesian personalized ranking \textbf{BPR}~\cite{rendle2009bpr}, the knowledge graph attention network \textbf{KGAT}~\cite{Wang2019KGATKG}, an adaptation of the \textbf{OpenDialKG}~\cite{moon2019opendialkg} DialKG Walker model, and an adaptation of \textbf{KBRD}~\cite{Chen2019KBRD}. For next-question prediction, we compare the popular response ranking methods  \textbf{DMN}~\cite{Yang2018ResponseRW}, \textbf{DAM}~\cite{Zhou2018MultiTurnRS}, and \textbf{MSN}~\cite{Zhou2018MultiTurnRS}.
We also invoked the adapted versions of OpenDialKG and KBRD on this task, where both of them exploit the knowledge graphs to better leverage sentence, dialogue, and KG structural features. We adopt pre-trained TransE \cite{bordes2013translating} as the encoding for each entity within the KG and word embeddings are trained using the word2vec~\cite{mikolov2013distributed} skip-gram model.
\subsection{Recommendation}
The recommendation quality results of different models are given in Table \ref{tab:eval}. Among the methods, BPR optimizes a pairwise ranking only considering user--product pairs, while KGAT integrates the knowledge graph reasoning for recommendation. The best results are obtained by our modified KBRD baseline.

\subsection{Next-Question Prediction}
At the same time, learning to ask an appropriate question is another important indicator of evaluating whether the model successfully identifies the user needs. Compared to generation-based methods \cite{Liu2019RhetoricallyCE, Chen2019TowardsKR}, retrieval-based methods are able to provide more fluent and informative responses \cite{Yang2018ResponseRW,Yuan2019MultihopSN}.
The question prediction in conversational recommendation seeks to better narrow down the user's needs and effectively retrieve the best-matching products. The experimental results are given in Table \ref{tab:res_next}.
OpenDialKG and KBRD obtain the best results here.
It should be noted that, due to computational resource constraints, how to fully utilize the unified KG structure to avoid comprehensive reasoning either based on semantic features of historical dialogue or the overall structure of the KG are key challenges.

\section{Conclusion}
We introduce the new \approach{} dataset for conversational knowledge-enhanced recommendation.
Our work is the first exploration of creating a conversational dataset for recommendation that simulates user feedback with regard to a knowledge graph.
Compared to previous work, it enables more realistic conversational recommendation as well as explainability.

In this work, we assume customers are rational and patient in their interactions with the intelligent agent.
In future work, we hope to introduce more challenging tasks, where the user is able to provide more diverse responses with richer semantics and varying sentiment towards product attributes, or present product-specific requests.

We make our data available at \url{https://github.com/zuohuif/COOKIE} with further updates and maintenance. Besides, we will also provide the results of baseline methods in order to support and encourage further research on conversational agents for e-commerce settings.

\bibliographystyle{acl_natbib}
\bibliography{paper,acl2020,anthology}

\begin{thebibliography}{25}
\expandafter\ifx\csname natexlab\endcsname\relax\def\natexlab#1{#1}\fi

\bibitem[{Bettman et~al.(1998)Bettman, Luce, and Payne}]{Bettman1998}
James~R. Bettman, Mary~Frances Luce, and John~W. Payne. 1998.
\newblock \href {https://doi.org/10.1086/209535} {{Constructive Consumer Choice
  Processes}}.
\newblock \emph{Journal of Consumer Research}, 25(3):187--217.

\bibitem[{Bordes et~al.(2013)Bordes, Usunier, Garcia-Duran, Weston, and
  Yakhnenko}]{bordes2013translating}
Antoine Bordes, Nicolas Usunier, Alberto Garcia-Duran, Jason Weston, and Oksana
  Yakhnenko. 2013.
\newblock Translating embeddings for modeling multi-relational data.
\newblock In \emph{Advances in neural information processing systems}, pages
  2787--2795.

\bibitem[{Chen et~al.(2019{\natexlab{a}})Chen, Lin, Zhang, Ding, Cen, Yang, and
  Tang}]{Chen2019TowardsKR}
Qibin Chen, Junyang Lin, Yichang Zhang, Ming Ding, Yukuo Cen, Hongxia Yang, and
  Jie Tang. 2019{\natexlab{a}}.
\newblock Towards knowledge-based recommender dialog system.
\newblock \emph{ArXiv}, abs/1908.05391.

\bibitem[{Chen et~al.(2019{\natexlab{b}})Chen, Lin, Zhang, Ding, Cen, Yang, and
  Tang}]{Chen2019KBRD}
Qibin Chen, Junyang Lin, Yichang Zhang, Ming Ding, Yukuo Cen, Hongxia Yang, and
  Jie Tang. 2019{\natexlab{b}}.
\newblock Towards knowledge-based recommender dialog system.
\newblock \emph{ArXiv}, abs/1908.05391.

\bibitem[{Greco et~al.(2017)Greco, Suglia, Basile, and
  Semeraro}]{Greco2017ConverseEtImperaED}
Claudio Greco, Alessandro Suglia, Pierpaolo Basile, and Giovanni Semeraro.
  2017.
\newblock Converse-et-impera: Exploiting deep learning and hierarchical
  reinforcement learning for conversational recommender systems.
\newblock In \emph{AI*IA}.

\bibitem[{Jannach et~al.(2020)Jannach, Manzoor, Cai, and
  Chen}]{jannach2020survey}
Dietmar Jannach, Ahtsham Manzoor, Wanling Cai, and Li~Chen. 2020.
\newblock A survey on conversational recommender systems.
\newblock \emph{arXiv preprint arXiv:2004.00646}.

\bibitem[{Lei et~al.(2020)Lei, He, Miao, Wu, Hong, Kan, and
  Chua}]{Lei2020EstimationActionReflectionTD}
Wenqiang Lei, Xiangnan He, Yisong Miao, Qingyun Wu, Richang Hong, Min-Yen Kan,
  and Tat-Seng Chua. 2020.
\newblock Estimation-action-reflection: Towards deep interaction between
  conversational and recommender systems.
\newblock \emph{Proceedings of the 13th International Conference on Web Search
  and Data Mining}.

\bibitem[{Li et~al.(2017)Li, Qiu, Chen, Wang, Gao, Huang, Ren, Zhao, Zhao,
  Wang, Jin, and Chu}]{Li2017AliMeA}
Feng-Lin Li, Minghui Qiu, Haiqing Chen, Xiongwei Wang, Xing Gao, Jun Huang,
  Juwei Ren, Zhongzhou Zhao, Weipeng Zhao, Lei Wang, Guwei Jin, and Wei Chu.
  2017.
\newblock Alime assist : An intelligent assistant for creating an innovative
  e-commerce experience.
\newblock \emph{Proceedings of the 2017 ACM on Conference on Information and
  Knowledge Management}.

\bibitem[{Li et~al.(2018)Li, Kahou, Schulz, Michalski, Charlin, and
  Pal}]{li2018conversational}
Raymond Li, Samira~Ebrahimi Kahou, Hannes Schulz, Vincent Michalski, Laurent
  Charlin, and Chris Pal. 2018.
\newblock Towards deep conversational recommendations.
\newblock In \emph{Advances in Neural Information Processing Systems 31 (NIPS
  2018)}.

\bibitem[{Liu et~al.(2019)Liu, Fu, Cao, de~Melo, Tam, Niu, and
  Zhou}]{Liu2019RhetoricallyCE}
Zhiqiang Liu, Zuohui Fu, Jie Cao, Gerard de~Melo, Yik-Cheung Tam, Cheng Niu,
  and Jie Zhou. 2019.
\newblock Rhetorically controlled encoder-decoder for modern chinese poetry
  generation.
\newblock In \emph{ACL}.

\bibitem[{Masson(1983)}]{Masson1983ConceptualPO}
Michael E.~J. Masson. 1983.
\newblock Conceptual processing of text during skimming and rapid sequential
  reading.
\newblock \emph{Memory \& Cognition}, 11:262--274.

\bibitem[{Mikolov et~al.(2013)Mikolov, Sutskever, Chen, Corrado, and
  Dean}]{mikolov2013distributed}
Tomas Mikolov, Ilya Sutskever, Kai Chen, Greg~S Corrado, and Jeff Dean. 2013.
\newblock Distributed representations of words and phrases and their
  compositionality.
\newblock In \emph{Advances in neural information processing systems}, pages
  3111--3119.

\bibitem[{Moon et~al.(2019)Moon, Shah, Kumar, and Subba}]{moon2019opendialkg}
Seungwhan Moon, Pararth Shah, Anuj Kumar, and Rajen Subba. 2019.
\newblock {OpenDialKG}: Explainable conversational reasoning with
  attention-based walks over knowledge graphs.
\newblock In \emph{Proceedings of the 57th Annual Meeting of the Association
  for Computational Linguistics}, pages 845--854.

\bibitem[{Ni et~al.(2019)Ni, Li, and McAuley}]{ni2019justifying}
Jianmo Ni, Jiacheng Li, and Julian McAuley. 2019.
\newblock Justifying recommendations using distantly-labeled reviews and
  fine-grained aspects.
\newblock In \emph{Proceedings of the 2019 Conference on Empirical Methods in
  Natural Language Processing and the 9th International Joint Conference on
  Natural Language Processing (EMNLP-IJCNLP)}, pages 188--197.

\bibitem[{Rendle et~al.(2009)Rendle, Freudenthaler, Gantner, and
  Schmidt-Thieme}]{rendle2009bpr}
Steffen Rendle, Christoph Freudenthaler, Zeno Gantner, and Lars Schmidt-Thieme.
  2009.
\newblock Bpr: Bayesian personalized ranking from implicit feedback.
\newblock In \emph{Proceedings of the 25th conference on uncertainty in
  artificial intelligence}, pages 452--461. AUAI Press.

\bibitem[{Sun and Zhang(2018)}]{Sun2018ConversationalRS}
Yueming Sun and Yi~Zhang. 2018.
\newblock Conversational recommender system.
\newblock In \emph{SIGIR '18}.

\bibitem[{Wang et~al.(2019)Wang, He, Cao, Liu, and Chua}]{Wang2019KGATKG}
Xiang Wang, Xiangnan He, Yixin Cao, Meng Liu, and Tat-Seng Chua. 2019.
\newblock Kgat: Knowledge graph attention network for recommendation.
\newblock In \emph{KDD}.

\bibitem[{Wiseman et~al.(2018)Wiseman, Shieber, and
  Rush}]{Wiseman2018LearningNT}
Sam Wiseman, Stuart~M. Shieber, and Alexander~M. Rush. 2018.
\newblock Learning neural templates for text generation.
\newblock In \emph{EMNLP}.

\bibitem[{Yang et~al.(2018)Yang, Qiu, Qu, Guo, Zhang, Croft, Huang, and
  Chen}]{Yang2018ResponseRW}
Liu Yang, Minghui Qiu, Chen Qu, Jiafeng Guo, Yongfeng Zhang, W.~Bruce Croft,
  Jun Huang, and Haiqing Chen. 2018.
\newblock Response ranking with deep matching networks and external knowledge
  in information-seeking conversation systems.
\newblock \emph{The 41st International ACM SIGIR Conference on Research \&
  Development in Information Retrieval}.

\bibitem[{Yuan et~al.(2019)Yuan, jie Zhou, Li, Lv, Zhu, Han, and
  Hu}]{Yuan2019MultihopSN}
Chunyuan Yuan, Wen jie Zhou, MingMing Li, Shangwen Lv, Fuqing Zhu, Jizhong Han,
  and Songlin Hu. 2019.
\newblock Multi-hop selector network for multi-turn response selection in
  retrieval-based chatbots.
\newblock In \emph{EMNLP/IJCNLP}.

\bibitem[{Zhang and Chen(2018)}]{Zhang2018ExplainableRA}
Yongfeng Zhang and Xu~Chen. 2018.
\newblock Explainable recommendation: A survey and new perspectives.
\newblock \emph{Found. Trends Inf. Retr.}, 14:1--101.

\bibitem[{Zhang et~al.(2018)Zhang, Chen, Ai, Yang, and
  Croft}]{zhang2018towards}
Yongfeng Zhang, Xu~Chen, Qingyao Ai, Liu Yang, and W~Bruce Croft. 2018.
\newblock Towards conversational search and recommendation: System ask, user
  respond.
\newblock In \emph{Proceedings of the 27th ACM International Conference on
  Information and Knowledge Management}, pages 177--186. ACM.

\bibitem[{Zhang et~al.(2014)Zhang, Lai, Zhang, Zhang, Liu, and
  Ma}]{zhang2014explicit}
Yongfeng Zhang, Guokun Lai, Min Zhang, Yi~Zhang, Yiqun Liu, and Shaoping Ma.
  2014.
\newblock Explicit factor models for explainable recommendation based on
  phrase-level sentiment analysis.
\newblock In \emph{Proceedings of the 37th international ACM SIGIR conference
  on Research \& development in information retrieval}, pages 83--92. ACM.

\bibitem[{Zhao et~al.(2019)Zhao, He, Yang, Dou, Huang, Ouyang, and
  Wen}]{zhao2019kb4rec}
Wayne~Xin Zhao, Gaole He, Kunlin Yang, Hongjian Dou, Jin Huang, Siqi Ouyang,
  and Ji-Rong Wen. 2019.
\newblock Kb4rec: A data set for linking knowledge bases with recommender
  systems.
\newblock \emph{Data Intelligence}, 1(2):121--136.

\bibitem[{Zhou et~al.(2018)Zhou, Li, Dong, Liu, Chen, Zhao, Yu, and
  Wu}]{Zhou2018MultiTurnRS}
Xiangyang Zhou, Lu~Li, Daxiang Dong, Yi~Liu, Ying Chen, Wayne~Xin Zhao, Dianhai
  Yu, and Hua Wu. 2018.
\newblock Multi-turn response selection for chatbots with deep attention
  matching network.
\newblock In \emph{ACL}.

\end{thebibliography}

\end{document}